\documentclass[twocolumn,aps,showpacs,preprintnumbers,amsmath,amssymb, superscriptaddress]{revtex4-1}
\usepackage{bm}
\usepackage{graphicx}
\usepackage{color}
\usepackage{xspace}
\definecolor{dblue}{rgb}{0,0,0.6}
\definecolor{dred}{rgb}{0.9,0,0}
\definecolor{dgreen}{rgb}{0,0.4,0}
\pdfoutput=1

\begin{document}

\title{Raman study of electron-phonon coupling in thin films of LiTi$_2$O$_4$ spinel oxide superconductor}

\author{D.~Chen}
\affiliation{Beijing National Laboratory for Condensed Matter Physics, and Institute of Physics, Chinese Academy of Sciences, Beijing 100190, China}
\affiliation{School of Physical Sciences, University of Chinese Academy of Sciences, Beijing 100049, China}
\author{Y.-L.~Jia}
\affiliation{Beijing National Laboratory for Condensed Matter Physics, and Institute of Physics, Chinese Academy of Sciences, Beijing 100190, China}
\affiliation{School of Physical Sciences, University of Chinese Academy of Sciences, Beijing 100049, China}
\author{T.-T.~Zhang}
\affiliation{Beijing National Laboratory for Condensed Matter Physics, and Institute of Physics, Chinese Academy of Sciences, Beijing 100190, China}
\affiliation{School of Physical Sciences, University of Chinese Academy of Sciences, Beijing 100049, China}
\author{Z.~Fang}
\affiliation{Beijing National Laboratory for Condensed Matter Physics, and Institute of Physics, Chinese Academy of Sciences, Beijing 100190, China}
\affiliation{School of Physical Sciences, University of Chinese Academy of Sciences, Beijing 100049, China}
\affiliation{Collaborative Innovation Center of Quantum Matter, Beijing, China}
\author{K.~Jin}
\affiliation{Beijing National Laboratory for Condensed Matter Physics, and Institute of Physics, Chinese Academy of Sciences, Beijing 100190, China}
\affiliation{School of Physical Sciences, University of Chinese Academy of Sciences, Beijing 100049, China}
\affiliation{Collaborative Innovation Center of Quantum Matter, Beijing, China}
\author{P.~Richard}\email{p.richard@iphy.ac.cn}
\affiliation{Beijing National Laboratory for Condensed Matter Physics, and Institute of Physics, Chinese Academy of Sciences, Beijing 100190, China}
\affiliation{School of Physical Sciences, University of Chinese Academy of Sciences, Beijing 100049, China}
\affiliation{Collaborative Innovation Center of Quantum Matter, Beijing, China}
\author{H.~Ding}\email{dingh@iphy.ac.cn}
\affiliation{Beijing National Laboratory for Condensed Matter Physics, and Institute of Physics, Chinese Academy of Sciences, Beijing 100190, China}
\affiliation{School of Physical Sciences, University of Chinese Academy of Sciences, Beijing 100049, China}
\affiliation{Collaborative Innovation Center of Quantum Matter, Beijing, China}

\date{\today}

\begin{abstract}
We performed a Raman scattering study of thin films of LiTi$_2$O$_4$ spinel oxide superconductor. We detected four out of five Raman active modes, with frequencies in good accordance with our first-principles calculations. Three T$_{2g}$ modes show a Fano lineshape from 5 K to 295 K, which suggests an electron-phonon coupling in LiTi$_2$O$_4$. Interestingly, the electron-phonon coupling shows an anomaly across the negative to positive magnetoresistance transition at 50 K, which may be due to the unset of other competing orders. The strength of the electron-phonon interaction estimated from the Allen's formula and the observed lineshape parameters suggests that the three T$_{2g}$ modes contribute little to superconductivity.
\end{abstract}



\maketitle

Because of their complexity of charge, magnetic and orbital degrees of freedom, transition-metal spinels have been studied intensely for many years~\cite{spinel}. As the only oxide superconductor with spinel structure, LiTi$_2$O$_4$ has aroused many researches since the discovery of its superconductivity \cite{firstSC}. This compound was initially compared to cuprates because both materials are transition-metal oxides, although the superconducting transition temperature $T_c \approx 12$ K of LiTi$_2$O$_4$ is not very high~\cite{calculation_cuprate_lamda,neutron_cuprate_DOS}. Unlike cuprates, both transport and spectroscopy results indicate that LiTi$_2$O$_4$ is a typical fully gapped type-II BCS superconductor~\cite{specificheat_BCS_lamda,criticalfield_BCS,Andreevreflection_BCS}. Later, with the development of lithium-ion batteries, Li$_{1+x}$Ti$_{2-x}$O$_4$ ($0 \leq x \leq \frac{1}{3}$) was studied intensively as an alternative electrode material~\cite{4512_electrochemicalLiinsertion,defectcalcul_battery,4512_usr_battery,ionicliquidgating}. 

Interest for LiTi$_2$O$_4$ was refreshed recently with the report of anomalous magnetoresistance at 50 K~\cite{magnetoresistance}, which suggests that spin-orbital fluctuations play an important role in LiTi$_2$O$_4$. Bosonic modes were also observed in tunneling spectra, reinforcing the assumption for an important electron-boson coupling~\cite{anisotropic_bosonicmode}. Although several works in the literature report the lattice dynamics of LiTi$_2$O$_4$, both theoretically and experimentally~\cite{Raman,phononcalcul1,A1gcalculation,phononcalcul2,phononcalcul3,phononcalcul4}, there is significant discrepancies among the various studies. Moreover, direct phonon measurements for coupling analysis have been limited because of the poor availability of single crystals~\cite{review}. In order to clarify the vibration modes and investigate the strength of the electron-phonon coupling, here we report a Raman scattering study of LiTi$_2$O$_4$ single-crystalline films supported by first-principles calculations. We detect four out of five Raman active modes, with frequencies in good accordance with our first-principles calculations. A Fano line shape was observed for the three T$_{2g}$ modes from 295 K to 5 K, which suggests that electron-phonon coupling is important in LiTi$_2$O$_4$. The electron-phonon coupling shows an anomaly around 50 K, where the negative to positive magnetoresistance occurs \cite{magnetoresistance}. The unset of other orders below 50 K, like orbital-related states, may quench the electron-phonon coupling, resulting in its fluctuation. Allen's formula was used to estimate the strength of the electron-phonon interaction from the observed lineshape parameters. Although they have asymmetric line shape, the three T$_{2g}$ modes contribute only little to the superconductivity, with an average electron-phonon coupling constant $\overline{\lambda}$ = 0.074.

\begin{figure}[!t]
\begin{center}
\includegraphics[width=\columnwidth]{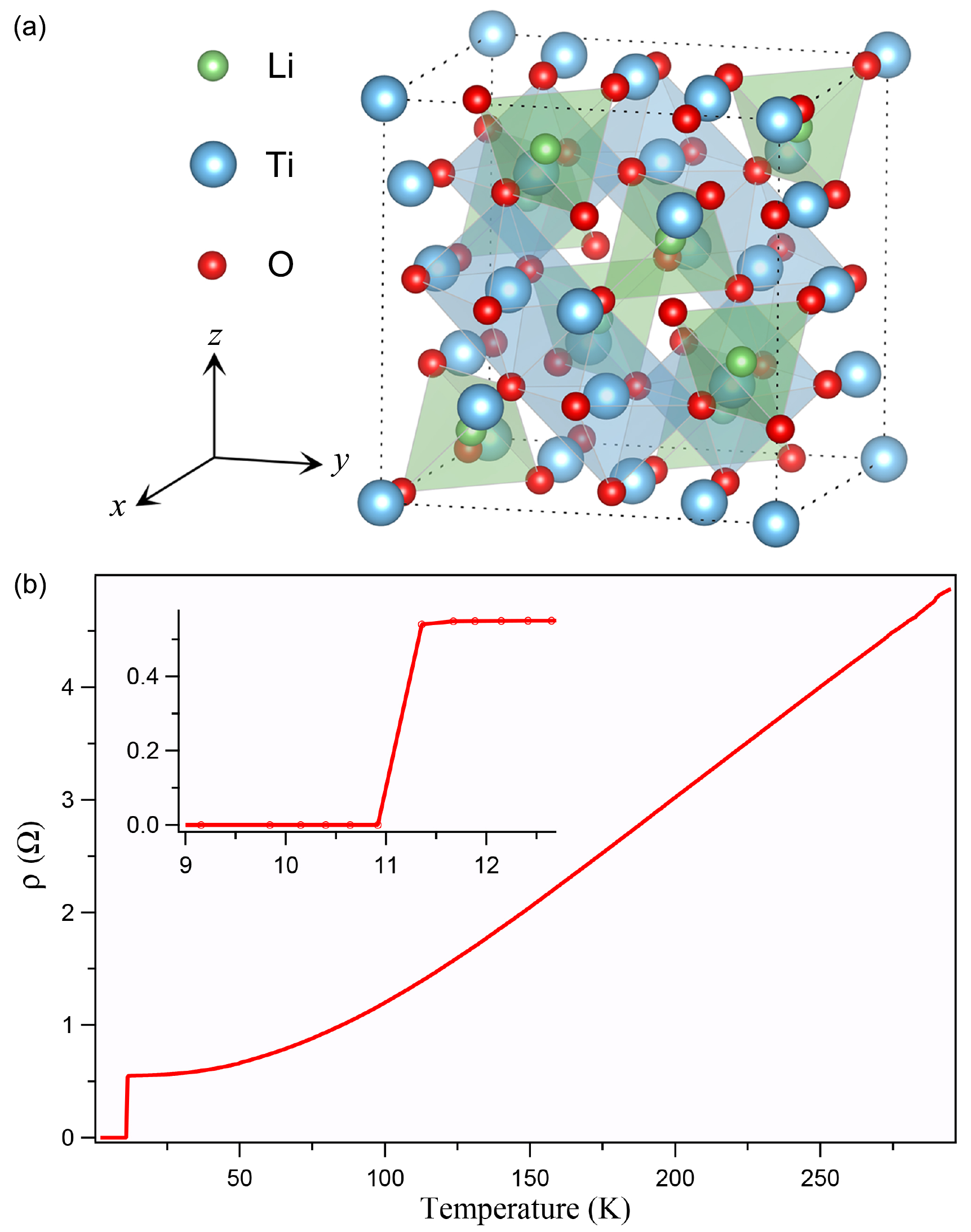}
\end{center}
\caption{\label{figure1}(Color online). (a) Crystal structure of LiTi$_2$O$_4$. Li atoms fill $\frac{1}{8}$ of the tetrahedral sites (in green) and Ti atoms fill $\frac{1}{2}$ of the octahedral sites (in blue). (b) In-plane resistivity of LiTi$_2$O$_4$. The inset is a close-up of the superconducting transition.}
\end{figure}

The LiTi$_2$O$_4$ films used in our Raman study were grown on MgAl$_2$O$_4$ (001) substrates by pulsed laser deposition. Films with thickness $\approx$ 200 nm have been characterized to be single-crystalline and in a pure phase \cite{transport}. The electrical resistivity of the films presented in Fig. \ref{figure1}(b) shows a sharp superconducting transition at $T_c$ $\approx$ 11.3 K. To avoid Li vacancies induced by contact to water in air \cite{water-oxidation,Lideficiency}, the films were measured in a ST500 (Janis) cryostat with a working vacuum better than $2\times 10^{-6}$ mbar. A long-focus distance $20\times$ objective was used for back-scattering micro-Raman measurements between 5 and 295 K. Low power 488.0~nm and 514.5~nm excitations from an Ar-Kr ion laser were used as incident light. The scattering light was analyzed by a Horiba Jobin Yvon T64000 spectrometer equipped with a nitrogen-cooled CCD camera. The confocal design of this spectrometer allows us to measure the signal from both films and substrates. Substrates without films were also measured as reference. As shown in Fig. \ref{figure1}(a), we define $x$, $y$ and $z$ as the directions along the unit cell axes. Raman spectra have been recorded under the $(\mathbf{\hat{e}}^i\mathbf{\hat{e}}^s)=$ ($xx$), ($yy$) and ($xy$) polarization configurations.

Space group $Fd\overline{3}m$ (point group $O_{h}$) characterizes the crystal structure of LiTi$_2$O$_4$ \cite{review}, which is presented in Fig. \ref{figure1}(a). A single unit cell contains two chemical formula units, for a total of 14 atoms. A simple group symmetry analysis \cite{bilbal} indicates that the phonon modes at the Brillouin zone (BZ) center $\Gamma$ decompose into [T$_{1u}$]+[4T$_{1u}$]+[A$_{1g}$+E$_{g}$+3T$_{2g}$]+[2A$_{2u}$+2E$_{u}$+2T$_{2u}$+T$_{1g}$], where the first, second, third and fourth terms represent the acoustic modes, the infrared-active modes, the Raman-active modes and the silent modes, respectively. To get estimates on the phonon frequencies, we performed first-principles calculations of the phonon modes at $\Gamma$ in the framework of the density functional perturbation theory (DFPT) \cite{DFPT2} without considering spin-orbit coupling. We adopted the fully-relaxed lattice parameters a = b = c = 8.4 \AA, and the Wyckoff positions (Li $8a$, Ti $16d$ and O $32e$) from experimental data~\cite{anisotropic_bosonicmode}. For all calculations, we used the Vienna \textit{ab-initio} simulation package (VASP) \cite{VASP2} with the generalized gradient approximation (GGA) of Perdew-Burke-Ernzerhof for the exchange-correlation functions~\cite{PBE}. The projector augmented wave (PAW) method \cite{PAW} was employed to describe the electron-ion interactions. A plane wave cut-off energy of 500 eV was used with a uniform $6\times6\times6$ Monkhorst-Pack $k$-point mesh for a $2\times2\times2$ supercell. The real-space force constants of the supercell were calculated using DFPT \cite{DFPT1} and the phonon frequencies were calculated from the force constants using the PHONOPY code \cite{PHONOPY}. The calculated optic mode frequencies, their symmetries and optical activities, as well as the main atoms involved, are given in Table \ref{EXP_CAL_comparsion}. Compared to previous calculation results with different methods, our results are more consistent with the experimental values.

\begin{table*}
\caption{\label{EXP_CAL_comparsion}Calculated optic mode frequencies and experimental Raman active modes at 294 K, along with previous experimental \cite{Raman} and calculation \cite{phononcalcul1,phononcalcul2,phononcalcul4} results.}
\begin{ruledtabular}
\begin{tabular}{ccccccccc}
 Sym. &	Activity  & Exp. (This work) & Exp. (Ref. \cite{Raman}) & Cal. (This work)	& Atoms involved & Ref. \cite{phononcalcul1} & Ref. \cite{phononcalcul2} & Ref. \cite{phononcalcul4}\\
\hline
T$_{2u}$& Silent &  &  &   116.0&  Ti, O & 128.5 & 165.0 & 141.2\\
E$_{u}$& Silent & &	   &  204.0&	Ti, O & 286.6 & 236.8 & 275.3\\
T$_{1u}$& I & &    &  279.5&	Li, Ti, O & 289.2 & 210.3 & 247.2\\
T$_{2u}$& Silent & &	 &    360.3&	Ti, O & 461.3 & 542.5 & 493.6\\
E$_{g}$& R & - & 200 &      366.0& 	O & 337.4 & 429.0 & 428.6\\
T$_{2g}$& R & 342.2 & 339 &      369.8&	Li, O & 288.7 & 344.2 & 196.8\\
T$_{1u}$& I & & &	 383.0& 	Li, Ti, O & 389.5 & 424.9 & 412.7\\
T$_{1g}$& Silent & & 	& 398.6& 	O & 397.9 & 429.0 & 413.0\\
T$_{1u}$& I & &   &   416.4&	Li, Ti, O & 506.8 & 508.6 & 515.3\\
T$_{2g}$& R & 433.3&  429 &   466.6&	Li, O & 516.8 & 542.4 & 343.2\\
E$_{u}$& Silent & &   &   469.4&	Ti, O & 565.4 & 603.3 & 581.0\\
A$_{2u}$& Silent & &   &   498.2&  Ti, O & 496.3 & 323.8 & 397.6\\
T$_{2g}$& R & 495.2&   494&   533.5&	Li, O & 687.7 & 652.4 & 596.8\\
T$_{1u}$& I & &     & 561.5& 	Li, Ti, O & 696.8 & 668.3 & 695.1\\
A$_{2u}$& Silent & &    &     593.7&	Ti, O & 650.5 & 664.6 & 659.2\\
A$_{1g}$& R & 625.4& 628&	 625.2&	O & 548.5 & 628.0 & 628.0\\

\end{tabular}
\end{ruledtabular}
\begin{raggedright}
I = infrared active, R = Raman active, Silent = not optically active. The unit of the numbers are ``cm$^{-1}$".\\
\end{raggedright}
\end{table*}

\begin{figure}[!t]
\begin{center}
\includegraphics[width=\columnwidth]{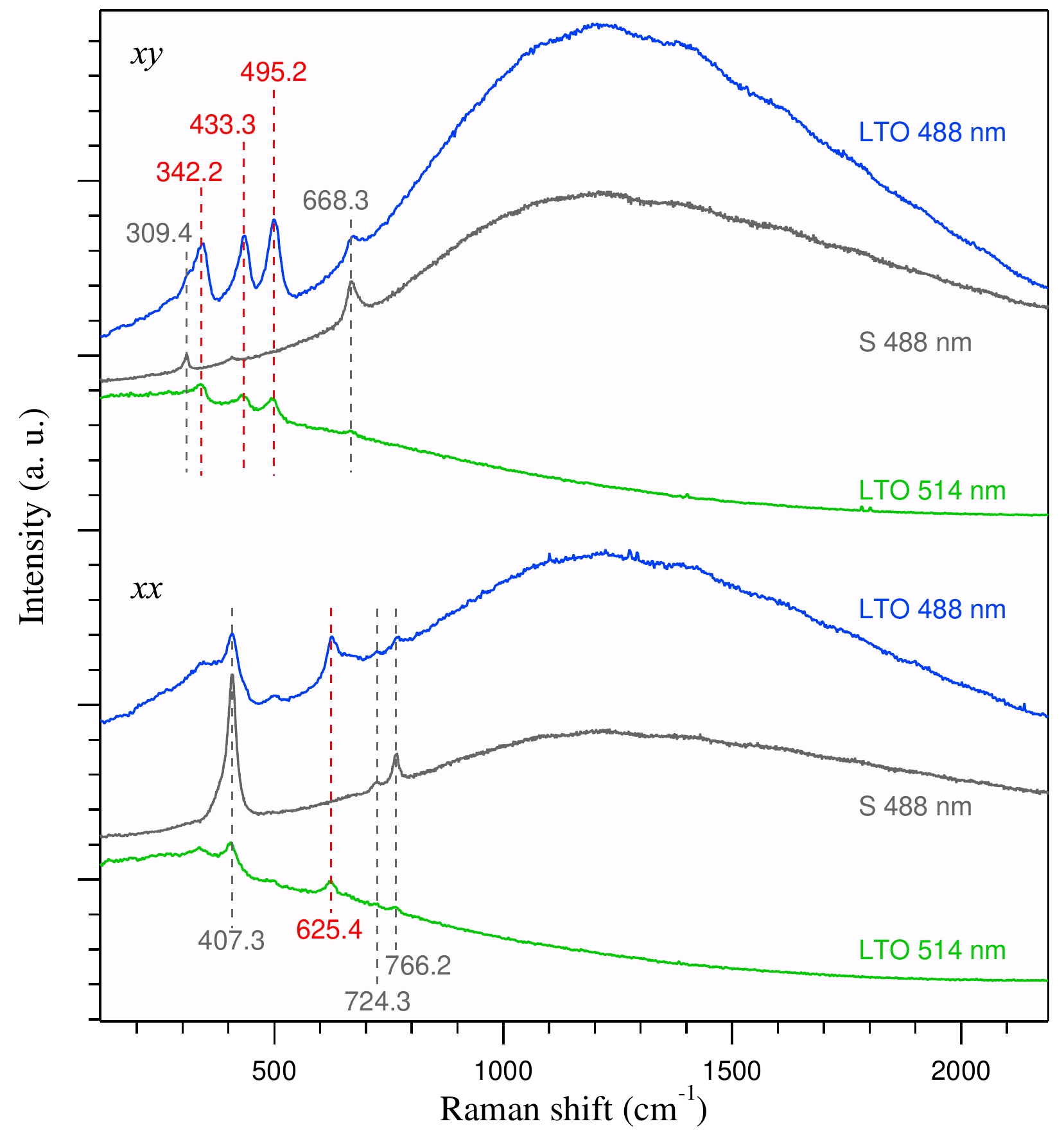}
\end{center}
\caption{\label{figure2}(Color online). Raman spectra of LiTi$_2$O$_4$ film and MgAl$_2$O$_4$ substrate recorded with 488.0~nm, 514.5~nm laser excitations under the ($xy$) and ($xx$) polarization configurations at room temperature. The red dashed lines and numbers indicate modes from the LiTi$_2$O$_4$ sample, while gray dashed lines and numbers indicate modes from substrate. ``LTO" and ``S" represent LiTi$_2$O$_4$ and substrate, respectively. The curves are shifted relative to each other for clarity.}
\end{figure}

In Fig. \ref{figure2}, we compare the Raman spectra of a LiTi$_2$O$_4$ film and of a substrate recorded at room temperature under different laser excitations. Based on literature, we can assign many peaks to the MgAl$_2$O$_4$ substrate. For instance, the peaks at 309.4~cm$^{-1}$ and 668.3~cm$^{-1}$ are T$_{2g}$ modes, the peak at 407.3~cm$^{-1}$ is an E$_{2g}$ mode, the one at 766.2~cm$^{-1}$ is an A$_{1g}$ mode, and the 724.3~cm$^{-1}$ excitation is from cation disorder~\cite{MgAlO4_4,MgAlO4_5,MgAlO4_6,MgAlO4_7,MgAlO4_9}. In addition to these peaks from the substrate, we observe 4 out of 5 Raman active modes predicted for LiTi$_2$O$_4$. The Raman tensors corresponding to the $O_{h}$ symmetry group are expressed in the $xyz$ coordinates as:
\begin{displaymath}
\textrm{A$_{1g}=$}
\left(\begin{array}{ccc}
a & 0 &0\\
0 & a &0\\
0 & 0 &a
\end{array}\right),
\end{displaymath}
\begin{displaymath}
\left[\begin{array}{cc}\textrm{E$_{g}=$}
\left(\begin{array}{ccc}
b & 0 &0\\
0 & b &0\\
0 & 0 &-2b\\
\end{array}\right),
\left(\begin{array}{ccc}
-\sqrt{3}b & 0 &0\\
0 & \sqrt{3}b &0\\
0 & 0 &0\\
\end{array}\right)
\end{array}\right],
\end{displaymath}
\begin{displaymath}
\left[\begin{array}{ccc}\textrm{T$_{2g}=$}
\left(\begin{array}{ccc}
0 & 0 &0\\
0 & 0 &d\\
0 & d &0\\
\end{array}\right),
\left(\begin{array}{ccc}
0 & 0 &d\\
0 & 0 &0\\
d & 0 &0\\
\end{array}\right),
\left(\begin{array}{ccc}
0 & d &0\\
d & 0 &0\\
0 & 0 &0\\
\end{array}\right)
\end{array}\right].
\end{displaymath}
\noindent Using the polarization selection rules, it is straightforward to assign the three peaks at 342.2~cm$^{-1}$, 433.3~cm$^{-1}$ and 495.2~cm$^{-1}$ to T$_{2g}$ modes, which are detected only under ($xy$) polarization configuration. As for the peak at 625.4~cm$^{-1}$ detected under ($xx$) polarization configuration, comparison with our calculation results suggests a A$_{1g}$ mode. The missing E$_{g}$ mode predicted at 366.0~cm$^{-1}$ may have too weak scattered intensity to be detected. Since its energy is close to one of the bosonic modes at around 40 meV observed in tunneling spectra \cite{anisotropic_bosonicmode}, it is possible that strong coupling reduces its lifetime. Apart from the phonon peaks, we observe a big hump at 1200~cm$^{-1}$ under 488.0~nm laser excitation in both the substrate and the LiTi$_2$O$_4$ thin film. Under 514.5~nm laser excitation, the hump shifts to 200~cm$^{-1}$. The humps correspond to the same transition at 520~nm (2.38~eV) that we assign to luminescence from the substrate.

\begin{figure}[!t]
\begin{center}
\includegraphics[width=\columnwidth]{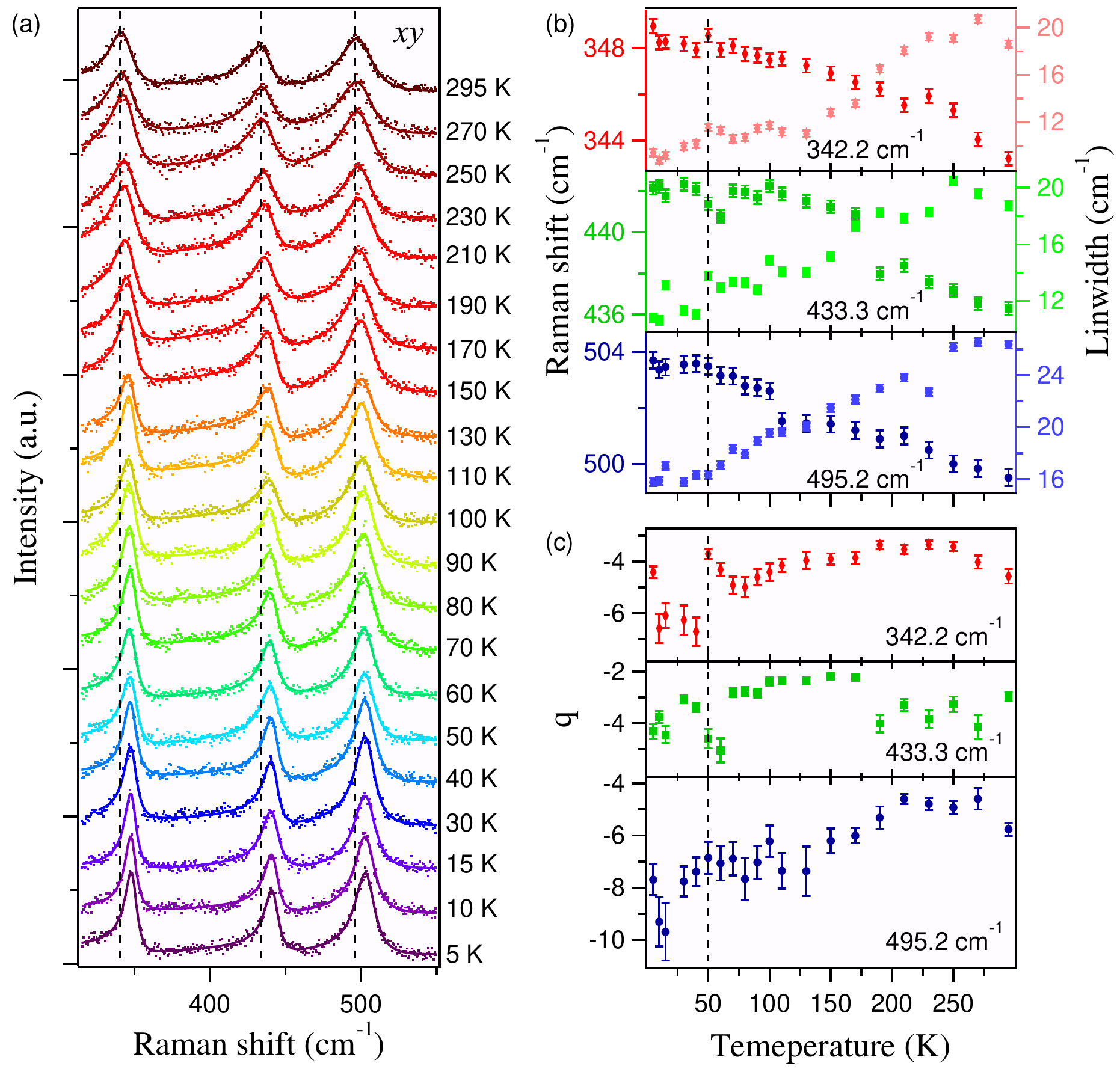}
\end{center}
\caption{\label{figure3}(Color online). (a) Waterfall plot of temperature dependent Raman spectra for the three T$_{2g}$ modes of LiTi$_2$O$_4$. The vertical dashed lines indicate these three modes. The colored dashed curves are data corrected by subtracting the substrate's data at the corresponding temperature. The colored curves are the resulting fitted spectra with Fano functions. (b) Temperature evolution of the renormalized phonon energies and linewidths. (c) Fano asymmetry parameters $q$ of the three T$_{2g}$ modes from the fits in (a). The vertical dashed line indicates 50 K. The error bars in (b) and (c) are from the system resolution and the fitting error, respectively.}
\end{figure}

We notice that the three T$_{2g}$ modes of LiTi$_2$O$_4$ have quite asymmetric line shapes, which implies a Fano resonance. The Fano resonance is a quantum interference between a discrete state and a continuum \cite{fano}. For Raman scattering, the spectrum of the phonon mode will present an asymmetric Fano line shape if there is an electronic-phonon coupling \cite{Ramanfano}. To further study the role of the electronic-phonon coupling in LiTi$_2$O$_4$, we analyzed the temperature dependence of the Raman spectra of the three T$_{2g}$ modes with Fano functions. As shown in Fig. \ref{figure3} (a), the Raman spectra can be well fitted by the equation:
\begin{equation}
\label{eq}
I(\omega) = \sum\limits_{i}\frac{A_{i}(q_{i}\Gamma_{i}/2+\omega-\omega_{i})^2}{(\Gamma_{i}/2)^{2}+(\omega-\omega_{i})^2}
\end{equation}
where $A_{i}$ is the amplitude, $\omega_{i}$ is the resonance energy (renormalized in the presence of the coupling), $\Gamma_{i}$ is the linewidth (full-width at half-maximum) and $q_{i}$ is the asymmetric parameter for the $i^{\textrm{th}}$ T$_{2g}$ mode. The factor $|1/q|$ is often used to estimate the electron-phonon coupling strength. The larger $|1/q|$, the stronger the coupling. The fitting results including the renormalized phonon energies, the linewidths and the Fano asymmetry parameters $q$ are displayed in Figs. \ref{figure3}(b) and \ref{figure3}(c). All the three T$_{2g}$ modes have higher energies and narrow linewidths upon cooling. Interestingly, there are anomalies around 50 K for the asymmetry parameters. This indicates that the electron-phonon coupling strength varies. We note that a magnetoresistivity transition from negative to positive is reported as temperature is decreased below 50 K, which suggests the presence of an orbital-related state \cite{magnetoresistance}. This unset of orbital order may quench the electron-phonon coupling, inducing the abnormal behavior of $|1/q|$.

We now follow a standard method to estimate the electron-phonon coupling strength associated to a particular mode $i$ using the Allen's formula \cite{Allen1,Allen4}:
\begin{equation}
\label{eq}
\lambda_{i} = \frac{2g_{i}\gamma_{i}}{\pi N_{\epsilon_{f}}\omega_{i}^2}
\end{equation}
where $\lambda_{i}$ is the dimensionless electron-phonon coupling constant, $g_{i}$ is the mode degeneracy, $N_{\epsilon_{f}}$ is the electronic density-of-states at the Fermi surface, $\omega_{i}$ is the mode energy and $\gamma_{i}$ is the linewidth. As an approximation at 5 K with $g_i$ = 3 and $N_{\epsilon_{f}}$ = 13.44 /eV unit cell \cite{magnetoresistance}, we get $\lambda_{1}$ = 0.089 for the T$_{2g}$(1) mode with $\omega_{1}$ = 348.9~cm$^{-1}$, $\gamma_{1}$ = 9.4~cm$^{-1}$, $\lambda_{2}$ = 0.063 for the T$_{2g}$(2) mode with $\omega_{2}$ = 442.2~cm$^{-1}$, $\gamma_{2}$ = 10.8~cm$^{-1}$, and $\lambda_{3}$ = 0.071 for the T$_{2g}$(3) mode with $\omega_{3}$ = 503.7~cm$^{-1}$, $\gamma_{3}$ = 15.8~cm$^{-1}$. The electron-phonon coupling constants for the three T$_{2g}$ modes are rather small with an average $\overline{\lambda}$ = 0.074, which is apparently inconsistent with a conventional phonon-mediated pairing mechanism. However, we caution that a more accurate evaluation of electron-phonon coupling constant would necessitate complete consideration of the contribution from all phonon modes across the entire first Brillouin zone.

In summary, we reported a polarized Raman scattering study of the only oxide spinel superconductor, LiTi$_2$O$_4$. Four out of five Raman active modes were detected, with frequencies in good accordance with our first-principles calculations. A Fano line shape was observed for the three T$_{2g}$ modes from 295 K to 5 K, which suggests that electron-phonon coupling is important in LiTi$_2$O$_4$. The electron-phonon coupling shows an anomaly around 50 K, where the negative to positive magnetoresistance occurs \cite{magnetoresistance}. The unset of other orders below 50 K, like orbital-related states, may quench the electron-phonon coupling, resulting in this anomaly. Allen's formula was used to estimate the strength of electron-phonon interaction from the observed lineshape parameters. Although they have electron-phonon coupling, the three T$_{2g}$ modes contribute little to the superconductivity.

We acknowledge G. He for useful discussions. This work was supported by grants from the National Natural Science Foundation (Nos. 11674371, 11274362, 11474338 and 11674374) and the Ministry of Science and Technology (Nos. 2015CB921301, 2016YFA0401000, 2016YFA0300300 and 2013CB921700) from China, and the Chinese Academy of Sciences (No. XDB07000000).

\bibliography{citation}

\end{document}